\documentclass[a4paper,fleqn,usenatbib]{mnras}

\usepackage{newtxtext,newtxmath}

\usepackage[T1]{fontenc}
\usepackage{ae,aecompl}

\usepackage{graphicx}	
\usepackage{amsmath}	
\usepackage{amssymb}	
\usepackage{soul}
\usepackage{ulem}

\title[Spin-down phase of neutron stars]{ACCRETION AND PROPELLER TORQUE IN THE SPIN-DOWN PHASE OF NEUTRON STARS: The case of transitional millisecond pulsar PSR J1023+0038}

\author[\"{U}. Ertan]{
\"{U}nal  Ertan,$^{1}$\thanks{E-mail: unal@sabaciuniv.edu}
\\
$^{1}$Sabanc\i\ University, 34956, Orhanl\i\, Tuzla, \.Istanbul,
Turkey\\
}

\date{Accepted XXX. Received YYY; in original form ZZZ}

\pubyear{2016}

\begin{document}
\label{firstpage}
\pagerange{\pageref{firstpage}--\pageref{lastpage}}
\maketitle

\def\be{\begin{equation}}
\def\ee{\end{equation}}
\def\ba{\begin{eqnarray}}
\def\ea{\end{eqnarray}}
\def\m{\mathrm}
\def\d{\partial}
\def\R{\right}
\def\L{\left}
\def\a{\alpha}
\def\acold{\alpha_\mathrm{cold}}
\def\ahot{\alpha_\mathrm{hot}}
\def\Mdotstar{\dot{M}_\ast}
\def\Omegastar{\Omega_\ast}
\def\Omegadot{\dot{\Omega}}

\def\OmegaK{\Omega_{\mathrm{K}}}
\def\Mdotin{\dot{M}_{\mathrm{in}}}
\def\Mdots{\dot{M}_{\mathrm{s}}}

\def\Mdotcrit{\dot{M}_{\mathrm{crit}}}
\def\Mdotout{\dot{M}_{\mathrm{out}}}

\def\Mdot{\dot{M}}
\def\Edot{\dot{E}}
\def\Pdot{\dot{P}}
\def\nudot{\dot{\nu}}
\def\Msun{M_{\odot}}

\def\Lin{L_{\mathrm{in}}}
\def\Lcool{L_{\mathrm{cool}}}
\def\Mdotstar{\dot{M}_\ast}
\def\Lstar{L_\ast}

\def\Rin{R_{\mathrm{in}}}
\def\rin{r_{\mathrm{in}}}
\def\rlc{r_{\mathrm{LC}}}
\def\rout{r_{\mathrm{out}}}
\def\rco{r_{\mathrm{co}}}
\def\re{r_{\mathrm{e}}}
\def\Ldisk{L_{\mathrm{disk}}}
\def\Lx{L_{\mathrm{x}}}
\def\Ld{L_{\mathrm{d}}}
\def\Lxpulsed{L_{\mathrm{x,pulsed}}}
\def\Md{M_{\mathrm{d}}} 
\def\NH{N_{\mathrm{H}}}
\def\dEb{\delta E_{\mathrm{burst}}}
\def\dEx{\delta E_{\mathrm{x}}}
\def\Bstar{B_\ast}\def\uff{\upsilon_{\mathrm{ff}}}
\def\Bb{\beta_{\mathrm{b}}}
\def\Be{B_{\mathrm{e}}}
\def\Bp{B_{\mathrm{p}}}
\def\Bz{B_{\mathrm{z}}}
\def\Bfi{B_{\mathrm{|phi}}}
\def\BA{B_{\mathrm{A}}}
\def\tint{t_{\mathrm{int}}}
\def\tdiff{t_{\mathrm{diff}}}
\def\r_m{r_{\mathrm{m}}}

\def\rA{r_{\mathrm{A}}}
\def\BA{B_{\mathrm{A}}}
\def\rS{r_{\mathrm{S}}}
\def\rp{r_{\mathrm{p}}}
\def\Tp{T_{\mathrm{p}}}
\def\dMin{\delta M_{\mathrm{in}}}
\def\Rc{\R_{\mathrm{c}}}
\def\Teff{T_{\mathrm{eff}}}
\def\uff{\upsilon_{\mathrm{ff}}}
\def\Tirr{T_{\mathrm{irr}}}
\def\Firr{F_{\mathrm{irr}}}
\def\Tcrit{T_{\mathrm{crit}}}
\def\P0min{P_{0,{\mathrm{min}}}}
\def\Av{A_{\mathrm{V}}}
\def\ah{\alpha_{\mathrm{hot}}}
\def\ac{\alpha_{\mathrm{cold}}}
\def\tc{\tau_{\mathrm{c}}}
\def\p{\propto}
\def\m{\mathrm}
\def\fast{\omega_{\ast}}
\def\Uff{\upsilon_{\mathrm{ff}}}
\def\Ufi{\upsilon_{\fi}}
\def\Ur{\upsilon_{\mathrm{r}}}
\def\UK{\upsilon_{\mathrm{K}}}
\def\Uesc{\upsilon_{\mathrm{esc}}}
\def\Uout{\upsilon_{\mathrm{out}}}
\def\Uphi{\upsilon_{\phi}}
\def\Udiff{\upsilon_{\mathrm{diff}}}
\def\Ure{\upsilon_{\mathrm{r,e}}}
\def\U{\upsilon}
\def\UB{\upsilon_{\mathrm{B}}}
\def\tauB{\tau_{\mathrm{B}}}
\def\hA{h_{\mathrm{A}}}
\def\he{h_{\mathrm{e}}}
\def\cs{c_{\mathrm{s}}}
\def\cse{c_{\mathrm{s,e}}}
\def\hin{h_{\mathrm{in}}}
\def\rhop{\rho^{\prime}}
\def\rhod{\rho_\mathrm{d}}
\def\rhos{\rho_\mathrm{s}}
\def\rhodp{\rho_\mathrm{d}^{\prime}}
\def\rhoe{\rho_\mathrm{e}}
\def\rhoout{\rho_\mathrm{out}}
\def\Alfven{Alfv$\acute{\mathrm{e}}$n~}
\def\418{SGR 0418+5729}
\def\142{AXP 0142+61}
\def\Caliskan{\c{C}al{\i}\c{s}kan~}
\def\ql{\textquotedblleft}
\def\qr{\textquotedblright~}

\def\gpers{g s$^{-1}$}
\def\ergpers{erg s$^{-1}$}
\def\Hzpers{Hz s$^{-1}$}
\def\spers{s s$^{-1}$}
\def\rinmax{r_{\mathrm{in,max}}}
\def\Rinmax{R_{\mathrm{in,max}}}

\begin{abstract}
The spin-down rate of  
PSR J1023+0038, one of the three confirmed transitional millisecond pulsars,  was measured in both radio pulsar (RMSP) and X-ray pulsar (LMXB) states.  The spin-down rate in the LMXB state is only about 27\% greater than in the RMSP state (Jaodand et al. 2016). The inner disk radius, $\rin$, obtained recently by Ertan (2017) for the propeller phase, which is close to the co-rotation radius, $\rco$, and insensitive to the mass-flow rate, can explain the observed torques together with the X-ray luminosities, $\Lx$ . The X-ray pulsar and radio pulsar states correspond to accretion with spin-down (weak propeller) and strong propeller situations respectively. Several times increase in the disk mass-flow rate takes the source from the strong propeller with a low $\Lx$ to the weak propeller with a higher $\Lx$ powered by  accretion on to the star. The resultant decrease in  $\rin$ increases the magnetic torque slightly, explaining  the observed small increase in the spin-down rate. We have found that the spin-up torque exerted by accreting material is much smaller than the  magnetic spin-down torque exerted by the disk in the LMXB state. \end{abstract}




\section{Introduction}

Observations of transitional millisecond pulsars (tMSPs) in their radio pulsar (RMSP) and X-ray pulsar (LMXB) states provide an excellent opportunity to test the torque and accretion luminosity models. At present, there are three confirmed tMSPs (Archibald et al. 2009, Papitto et al. 2013,  Bassa et al. 2014). These sources undergo occasional transitions between LMXB and RMSP states within about days to weeks timescales (Papitto et al. 2013, Stappers et al. 2014), and remain in one or the other state for months to years, much longer than dynamical or viscous timescales (see e.g. Jaodand et al. 2016). 

The three tMSPs,  namely PSR J1023+0038, XSS J12270--4859, and IGR J18245--2452 show remarkably similar behaviours.  In the  RMSP state, sources show radio pulses and orbitally modulated X-ray emission. In the LMXB state, radio pulses disappear, and the systems exhibit three well-defined, characteristic modes.  Coherent X-ray pulsations are observed only in the high mode with an X-ray luminosity $\Lx \sim$   a few $ \times 10^{33}$ \ergpers, which is $\sim 5 - 7$ times greater than $\Lx$ in the low mode. In the LMXB state, the sources remain in the high mode for about $70 - 80 \%$ of the time, and in the low mode in the remaining time. The stars also exhibit occasional short X-ray flares with luminosities of $\Lx \sim 5 \times 10^{34}$ \ergpers (Linares 2014, Papitto et al. 2015, Jaodand et al. 2016).  
The most likely origin of the pulsed X-ray luminosities of XSS J12270$-$4859 and PSR J1023+0038  (hereafter  J1023) seems to be the  mass flow onto the neutron star channeled by the field lines (Papitto et al. 2015, Archibald et al. 2015). All three sources are spinning down with $\Pdot \simeq 6.83\times 10^{-20}$ s s$^{-1}$  for J1023 (Archibald et al. 2013 ), $\Pdot \simeq 1.11\times 10^{-20}$ s s$^{-1}$ for  XSS J12270--4859 (Ray et al. 2015),  and $\Pdot < 1.3 \times 10^{-17}$ s s$^{-1}$ for IGR J18245--2452 (Papitto et al. 2013) in the RMSP state. A measurement of $\Pdot \simeq 8.7\times 10^{-21}$ s s$^{-1}$ has been reported  for the high mode of the LMXB state of J1023 (Jaodand 2016).

These discoveries are rather surprising, because, according to conventional models, neutron stars are expected to be in the propeller phase without any accretion for  low X-ray luminosities (Illarionov \& Sunyaev 1975). In many theoretical models,  inner radius of the disk, $\rin$, is estimated to be close to the conventional \Alfven radius, $r_A$, while mass accretion onto the star is expected when the innermost disk extends inward of  the co-rotation radius, $\rco$, at which the speed of field lines co-rotating with the star equals the Kepler speed of the disk matter. This is in sharp contrast with the properties of tMSPs  observed in quiescence at low X-ray luminosities.  For  J1023, the high mode with pulsed X-rays suggests accretion with spin down, in a weak propeller state with $\rin > \rco$, while  
$\rA$  is estimated to be about 6 times greater than $\rco$. Such an $\rA$ has no physical meaning since it remains even outside the light cylinder. This clearly indicates that the actual $\rin$ in the spin-down phase could be much smaller than  $\rA$.         
 
As shown recently (Ertan 2017) the maximum $\rin$ at which the propeller mechanism can work is much smaller than  $\rA$, but not much larger than $\rco$.  
The critical accretion rate, $\Mdotcrit$ for the transition to accretion with spin-down (weak propeller)  is orders of magnitude smaller than the rate corresponding to $\rA \simeq \rco$. this is consistent with the  luminosities of tMSPs during the transitions between the RMSP and LMXB states. 
In the propeller phase,  the disk mass-flow rate, $\Mdotin$,  dependence of $\rin$ is much weaker than that of $\rA$, and variations in
magnetic torque in response to changes in $\Mdotin$ are much smaller than in conventional torque models with $\rin \p  \rA$.   These predictions of the model can be tested with precisely determined properties of J1023 in different states of the spin-down phase.  
    
We pursue these results obtained by Ertan (2017) to model the spin-down torques and the luminosities of the tMSPs. 
Our model is described in Section 2. In Section 3, we test the model with the rotational properties and X-ray luminosities of   J1023 in LMXB and RMSP states; 
and estimate the dipole field strengths and $\Mdotcrit$ for the MSPs in compact LMXBs, also known as \lq redbacks\rq~  (see Linares 2014 for a review) that have been observed only in the RMSP state.  We discuss our results in Section 4.     
               
\section{The Spin-down Phase}

The magnetosphere of the star is defined as the region of closed field lines in which matter and the field lines rotate together. The inner disk radius, $\rin$, is expected to be close to the radius of the magnetosphere. The inner disk and the field lines interact in a boundary between $\rin$ and $\rin + \Delta r$. In the interaction region,  
the field lines cannot slip through the disk, because the diffusion timescale of the magnetic field lines  (which is comparable to the viscous timescale, $t_\m{visc}$) is much longer than the the interaction timescale  $\tint \simeq |\Omegastar - \OmegaK|^{-1}$  (Fromang \& Stone 2009) where $\Omegastar$ is the rotational angular velocity of the star, and  $\OmegaK $ is the Keplerian angular velocity of the disk matter.   
The field lines interacting with the disk inflate and open up on the interaction timescale  (Aly 1985, Lovelace et al. 1995, Hayashi et al. 1996, Miller \& Stone 1997, Uzdensky et al. 2002, Uzdensky 2004).  
In the propeller phase, the matter leaves the disk along open field lines, and  the field lines reconnect  on dynamical timescale  completing the cycle  (Lovelace et al. 1999, Ustyugova et al. 2006). This could be imagined as a continuous process since $t_\m{dyn} \ll t_\m{visc}$. Numerical simulations  indicate that the field lines outside a radially narrow boundary remain disconnected from the disk (Lovelace et al. 1995). 

The critical condition for a steady propeller effect is defined by  first principles: (i) at a radius greater than $r_1 = 1.26 ~\rco$ the speeds of the field lines co-rotating with the star exceed the escape speed, $\Uesc$.  (ii) the angular momentum  transferred to gas at $\sim \rin > r_1$ should  accelerate the matter to the speed of the field lines within $\tint$.  The maximum inner disk radius, $\rinmax$, at which the propeller condition is satisfied is estimated as
\be
\Rinmax^{25/8} ~(1 -  \Rinmax^{-3/2})
~\simeq ~8.4  ~
\a_{-1}^{2/5} ~M_1^{-7/6} ~\Mdot_{12}^{-7/20}~ \mu_{26} ~ P_{-3}^{-13/12} 
\label{103}
\ee  
(Ertan 2017), where $\Rinmax = \rinmax / \rco$,  $\a_1 = (\a /0.1)$,  $M_1  = (M / \Msun)$,  $\Mdot_{12} = \Mdotin / (10^{12} $ \gpers), $\mu_{26} = \mu / (10^{26} $ G cm$^3$), and $P_{-3}$ is the rotational period of the star in milliseconds. We have obtained equation (\ref{103}) rearranging equation (9) in Ertan (2017) which gives $\rinmax$ in terms of $\rA$.
We write  $\rin = \eta ~\rinmax$ with $\eta \lesssim 1$  likely to be close to unity as magnetic stresses decrease sharply with r.
 The  $\Mdotin$ dependence of  $\rin$ is much weaker than that of $\rA \p \Mdot^{-2/7}$. 

The radial width of the outflow region, $\delta r$, is likely to be very narrow. Estimated as the radial diffusion length of the inner disk within several $\tint$, it is found to be even smaller than the pressure scale-hight of the disk $h$. 
The interaction does not have to take place only in such a tiny boundary  where the propeller condition is satisfied. While matter can leave the system from a radially very thin region of width  $\delta r$ at $\rin$, matter expelled by field lines in a wider region,  between $\rin$ and $\rin + \Delta r$, could return back to disk  at larger radii. We will use the term \lq backflow\rq ~for matter that is propelled from the boundary and  
returns back to the disk, and reserve \lq outflow\rq ~for the matter leaving the system from $\rin < r < \rin + \delta r$ with speeds greater than $\Uesc$.  A steady propeller state is reached, as shown in Fig. 1,  when the net mass-flow rate at each point along the disk becomes equal to $\Mdotin$. When  there is a continuous mass backflow from the boundary to outer radii, the total flow along  the disk is inwards. Despite the pile-up resulting from the backflow, with $\Mdot_\m{total} = \Mdot_\m{back}+ \Mdotin$, $\rin$ remains constant because of efficient outflow from $\rin > r_1$ with $\Mdotout = \Mdotin$. 

For a constant $\Mdotin$, what happens if $\rin$ is instantaneously set up between $\rco$ and $r_1$?  There could be an efficient backflow from the boundary. 
Without any accretion or outflow, pile-up outside $\rin$ grows up in time, and pushes $\rin$ toward $\rco$. 
Backflowing matter from the boundary to larger radii  of the disk moves inwards and piles up outside $\rin$ on the viscous timescale across this region. This could take a long time due to the low $\Mdotin$. Furthermore, both $\tint$ and the field strength increase as $\rin$ decreases toward $\rco$, which requires more pile-up to push $\rin$ further inward. It is likely that $\rin$ decreases very slowly for a steady and low $\Mdotin$.    
This problem could be studied through numerical simulations. In this work, we assume that a long-lasting propeller state could be maintained when $\rco < \rin < r_1$, as well as when $r_1 < \rin < \rinmax$.     

For weak-propeller state, we estimate that  the inner disk cannot penetrate inside $\rco$, and 
$\rin$ remains equal to  $\rco$ even when $\Mdotin$ is much greater than  $\Mdotcrit$ (but smaller than the conventional transition rate corresponding to $\rA \simeq \rco$, for the start for the spin-up phase), as long as $\rA > \rin \simeq \rco$.
This is because $\tint$ increases as $\rin$ approaches $\rco$, and
the gas co-rotating with the field lines, but not reaching $\Uesc$, can flow onto the star coupling to the field lines at $\rco$. The only way for $\rin$ to penetrate inside $\rco$ is that the viscous stresses should dominate the magnetic stresses at $\rco$, which is possible when $\rA$ comes close to $\rco$.  This will happen at accretion rates much larger than  the rates estimated for J1023 in its  LMXB state (Ertan 2017). 
Such high accretion rates $\Mdotin \sim 10^{17}$ \gpers are typical throughout the evolutionary epoch when accretion is spinning the neutron star up towards millisecond periods (Alpar et al. 1982, Radhakrishnan and Srinivasan 1982). From an evolutionary point of view, tMSPs are at the end of the LMXB epoch, with $\Mdotin$  from their companions much reduced as they proceed through the transition to the RMSP epoch, while they could still show X-ray outbursts  due to viscous disk instabilities like observed in IGR J18245--2452 (Linares 2014). 

In the weak-propeller phase of a tMSP, we can safely take  $\rin = \rco$  when $\eta ~\rinmax$  becomes smaller than $\rco$, and write 
\be
\rin = \m{max} \{\eta ~ \rinmax, \rco\}. 
\label{104}
\ee 
For a narrow boundary width $\Delta r$, and a mass accretion rate $\Mdotstar$, the total torque acting on the star becomes  
\be
\Gamma = - ~\frac {\mu^2}{\rin^3} \left(\frac{\Delta r}{\rin}\right) ~+~ (G M \rco)^{1/2} ~\Mdotstar
\label{105}
\ee 
where $G$ is the gravitational constant, and we take $\Mdotstar = \Mdotin$ in the accretion phase ($\rin = \rco$), and $\Mdotstar = 0$ in the strong propeller phase ($\rin > \rco$).  The first term on the right-hand side of equation (\ref{105}) is the spin-down torque resulting from the disk-field interaction, while the second-term is the spin-up torque due to angular momentum transfer by matter flowing onto the star. In the weak-propeller phase, the spin-down torque dominates the spin-up torque. Note that the spin-down torque in equation (\ref{105}) also includes and is much greater than  angular momentum loss through mass outflow  (see Ertan 2017).  

The X-ray luminosity powered by accretion onto the star is given by  
\be
\Lstar =  ~\frac {G M \Mdotstar}{R} 
\label{106}
\ee 
where $R$ is the radius of the star. In the strong propeller phase, the X-rays produced by viscous heating in the disk are emitted mostly from the  inner disk with luminosity 
 \be
\Ld =  ~\frac {G M \Mdotin}{2 ~\rin} 
\label{107}
\ee  
(see e.g. Frank et al. 2002) where $\rin = \eta ~\rinmax > \rco$ in this phase. In our model for J1023, $\Lstar$ with $\Mdotstar = \Mdotin$, the weak propeller,  represents the observed $\Lx$ in the high X-ray mode, while $\Lx = \Ld$ in the low X-ray mode (strong propeller) with negligible accretion onto the star.  We estimate  that  $\Mdotin$ values for these two modes in the LMXB state are similar. In the RMSP state,  $\Lx = \Ld$ like in the low mode, but with several times lower $\Mdotin$.

\begin{figure}
\vspace{-2.0cm}
\hspace{-2.0cm}
\includegraphics[width=0.7\textwidth=0.0,angle=0]{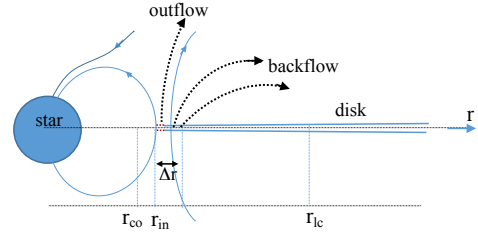}
\vspace{-12.5cm}
\caption{A simplified picture for the strong propeller phase.  When $\rin > 1.26  \rco$ a steady propeller phase could be built up. A long-lasting propeller phase could prevail when $\rco < \rin < 1.26 \rco$,  since $\rin$ is very insensitive to $\Mdotin$ and mass accumulation. In this case (with only back-flow), $\rin$ moves slowly inward due to increasing pile-up at the inner disk. Accretion (weak propeller) starts when $\rin \simeq \rco$,  and  the accretion torque is negligible compared to the spin-down torque (see the text).}
\end{figure}

Recently, simultaneous X-ray and radio continuum observations of J1023 in the LMXB state clearly showed an anti-correlation between the radio brightness and X-ray luminosity (Bogdanov et al. 2018). The radio flux, which is relatively steady in the high mode, starts to increase with transition to the low mode, reaches a maximum, and decreases to the pre-transition level within about 30 s of the transition back to the high mode. This picture is consistent with transition from the weak propeller to the strong propeller on short viscous timescale of the innermost disk. These transitions could take place with small occasional enhancements in the mass-outflow rate which could hinder the accretion on to the star briefly. The resultant density gradients at the inner disk lead to a rapid increase in $\Mdotin$, back to pre-enhancement level. Eventually, the accretion resumes switching on the high mode. The observed increase (decrease) in the radio brightness during the low mode is likely to be associated with the increasing (decreasing) rate of mass outflow, which is likely source of the unpulsed radio continuum emission. 

\section{Application to PSR J1023+0038}

The X-ray flux of  J1023 has been measured several times with {\it XMM-Newton} in the LMXB state in both low and high X-ray modes  (Archibald et al. 2015, Bogdanov et al. 2015, Jaodand et al. 2016). For the estimated distance $d \simeq 1365$ pc  (Deller at al. 2012),  $\Lx \sim 3\times 10^{33}$ \ergpers ~in the high mode, and  $\sim (3 - 5) \times 10^{32}$ \ergpers ~ in the low mode. In the RMSP state, $\Lx \sim 1 \times 10^{32}$ \ergpers (Archibald et al. 2010). The X-ray emission in the RMSP state is modulated with the orbital period (see the discussion below). Characteristic properties of  J1023  and other tMSPs in different states  (Linares 2014, Papitto et al. 2015, Archibald et al. 2015, Jaodand et al. 2016)  can be studied in the propeller model at hand. 

The period, $P$, and the period derivative, $\Pdot$, of J1023 were determined  in the RMSP phase  (Archibald et al. 2013 ). 
Recently  measured $\Pdot$  in the high X-ray mode shows that  $\Pdot$ increased by  $\sim$ 27 \% in the LMXB state compared to the torque measured earlier in the RMSP state (Jaodand et al. 2016).  

To explain observed changes in the $\Pdot$ and $\Lx$ of J1023, 
we propose: (1) in the high X-ray mode, the source is in the weak-propeller phase   with $\rin = \rco$, and $\Lx$ is produced by the accretion to the neutron star  surface with $\Mdotin = \Mdotstar$, (2) $\Lx$ in the low X-ray mode is produced by the inner disk with $\rin \simeq \rco$ with the same $\Mdotin$ as in the accretion phase, while $\Mdotstar = 0$, (3) in the RMSP state, $\Mdotin$ is several times less  than in the LMXB state, and the source is in the strong  propeller phase with $\Lx = \Ld$  and $\rin > \rco$. We do not address the sporadic X-ray flares observed from the source for about 2 \% of the time (see e.g. Jaodand et al. 2016 for a review of proposed models).    

For the torque calculation, we assume that $\Delta r / \rin$ is independent of $\Mdotin$ and $\rin$,  and use the same $\Delta r / \rin$ in the LMXB and RMSP states. For the transition from the LMXB to RMSP phase, we only decrease $\Mdotin$ by a factor which produces the observed $\Lx \simeq \Ld$ in the RMSP state. The resultant increase in $\rin$ is the main reason for the  decrease in the magnitude of the spin-down torque. 

In the LMXB phase, since $\Mdotin$ is very close to the transition rate,  the source could make occasional transitions between the weak-propeller and strong-propeller phases while $\rin \simeq \rco$. This explains  the high and low mode luminosities. 
In the low mode, a small fraction of $\Mdotin$ could still be flowing onto the star preventing the pulsed radio emission.  
In the RMSP state, since $\rin > \rco$, the accretion is not allowed,  and $\Lx = \Ld$ decreases by a factor  $2 \rin /R$ in comparison with $\Lx = \Lstar$ in the weak-propeller phase. 
\begin{table*} 
\caption{The model parameters and results for J1023. These results are obtained with $\eta = 0.78$, $\Delta r /\rin =0.1$, and the magnetic dipole moment $\mu_{26} =0.5$. The $L_\m{x,model}$ values for the high mode and the low mode corresponds to $\Lstar$   and $\Ld$ respectively. In the last column, $\Gamma_\m{acc}$  and $\Gamma_\m{tot}$ are the magnitudes of the accretion (spin-up) and  the spin-down torques respectively. }

\begin{tabular}{l|c|c|c|c|}
\hline 
& \multicolumn{2}{c|}{ LMXB state }&& RMSP state \\
\cline{2-3} \cline{5-5}
& High mode &Low mode && \\
 \hline
 \hline 
$L_\m{x,obs}~(10^{33}$\ergpers)   & 2.94 - 3.31   &0.31 - 0.55&&   0.094(6)\\
$L_\m{x,model}~(10^{33}$\ergpers)  &   2.97  & 0.62   && 0.10\\
\hline
 $\dot{P}_\m{obs}$~(\spers)   &$8.665\times10^{-21}$& -- &&  $6.834\times10^{-21}$\\
$\dot{P}_\m{model}$~(\spers)  & $8.72\times10^{-21}$& $8.72\times10^{-21}$&& 
$6.82\times10^{-21}$\\
\hline
$\dot{M}_\m{in}$~($10^{13}$~\gpers) &   1.6  & 1.6&&0.28\\
 $\dot{M}_{\ast}$ ~($10^{13}$~\gpers) & 1.6  & --   &&--\\
 $\rin / \rco$& 1  & 1&&1.09\\
 $\rin / \rA$& 0.16  & 0.16&&0.11\\
 $\Gamma_\m{acc} /\Gamma_\m{tot}$& $1.8\times10^{-2}$  & 0 && 0\\
 \hline
\end{tabular}
\end{table*}

\begin{table*}
\caption{The $\Mdotin$ and $\mu_{26}$ values that can produce the observed $P$, $\Pdot$, and $\Lx$ in the model for the three redback sources ($^{a}$Crawford et al. 2013, $^{b}$Bogdanov et al. 2014, $^{c}$Abdo et al. 2013, $^{d}$Gentile et al. 2014, $^{e}$Ray at al. 2015, $^{f}$Pletsch \& Clark 2015, $^{g}$Romani \& Shaw 2011, $^{h}$Linares 2014). }

\begin{tabular}{ c|c|c|c|c|c|c|c}
\cline{1-8}
& \multicolumn{3}{c|}{ Observed}& &\multicolumn{3}{c}{ Model} \\\cline{2-4}\cline{6-8}
 &$P$  & $\dot{P}_\m{obs}$~&$\Lx$ &
 &$\Mdotin$& $\Mdotcrit / \Mdotin$&$\mu_{26}$\\
 &(ms)&($10^{-20}$ \spers)&$(10^{32}$ \ergpers) & &($10^{12}$ \gpers)& & \\
 \hline
 \hline
 J1723~  & 1.855$^{a}$   & 0.75$^{a}$& 2.4$^{b}$ & &7.0 & $\lesssim 2$&$\sim $ 0.5\\
\hline
 J2215   &2.609$^{c}$& 3.3$^{c}$& 1.3$^{d, h}$ && 5.0 & $\sim 3$&$\sim $ 1.2  \\
 \hline
J2339  & 2.884$^{e}$& 1.41$^{f}$ & 2.7$^{g, h}$ && 1.1 &$\sim 7 $ &$\sim $ 0.8 \\
\hline
\end{tabular}
\end{table*}

Model parameters are compared  with observational properties of  J1023  in Table 1. These results are obtained with $\eta = 0.78$, $\Delta r /\rin =0.1$, and the magnetic dipole moment $\mu_{26} =0.5$. It is seen that the model can produce the $\Pdot$ values measured in the RMSP and LMXB states, consistently with the observed X-ray luminosities.    
The $\Mdotin$ in the LMXB states is about 6 times greater than in the RMSP state. This changes $\rin$  only by a small factor of $\sim 1.09$  due to very weak dependence of $\rin$ on $\Mdotin$ . The resultant increase in the spin-down torque is also small ($\sim 27 \%$)  in good agreement with the observations. Since the 
 accretion and dipole torques are only a few per cent of the torque produced by the disk-field interaction, the onset of accretion with the transition to the weak-propeller phase does not  affect $\Pdot$ significantly.     With $\mu_{26} = 4. 7$,  estimated in our model, the dipole torque is found to be about 5 and 7 times smaller than the observed torques in the RMSP and LMXB states respectively.
We note that for $\rin$ scaling with  $\rA$ in the propeller phase, the magnetic torque would be proportional to $\Mdotin^{6/7}$, implying changes by a factor $\sim 5$, not in agreement with the measured torques.

For the RMSP state, the observed $\Lx \sim 10^{32}$ \ergpers given in Table 1 is the 0.5 - 10 keV luminosity estimated by Archibald et al. (2010). From the spectral fits, Li et al. (2014) estimated that the 3 -79 keV luminosity is $ ~5 \times 10^{32}$ \ergpers. In both observations, the X-ray flux was found to be modulated with the orbital motion, which is likely to be due to an intra-binary shock produced by the interaction of pulsar wind with the matter outflowing from the companion.  A shock region close to the inner Lagrangian point that is eclipsed when the companion star is between the Earth and the neutron star could explain the observed modulations  (Archibald et al. 2010, Bogdanov et al. 2011,  see also Li et al. 2014 for a different interpretation). From the orbital modulations, it is expected that at least half of the X-rays are produced at the shock region in the RMSP state. These observations do not exclude a continuous emission from the disk with $\rin$ close to $\rco$ characterised by the conditions around the inner disk in the RMSP state. A large fraction of the total $\Lx$ is emitted by the intra-binary shock modulated by the orbital motion, while the remaining smaller fraction could be produced by the inner disk. A self-consistent explanation of the observed torque by the disk torques requires that $\Lx \sim 10^{32}$  \ergpers. This is  compatible with (smaller than) the continuous portion of the total X-ray luminosity in the RMSP state. We note that the disk emission spectrum could significantly change in the RMSP state due to strong propeller mechanism which can produce not only mass outflow but also hot matter around the inner disk.

The other two tMSPs, XSS J12270--4859 and IGR J18245--2452, also show X-ray modes similar to those of  J1023. For  XSS J12270--4859, $P = 1.69$ ms and $\Pdot = 1.11\times 10^{-20}$ s s$^{-1}$ (Ray et al. 2015). In the RMSP state, there is only an upper limit to the pulsed X-ray luminosity ($\Lxpulsed < 1.6\times 10^{31}$ \ergpers). For $\mu_{26} = 0.5$, we estimate $\Mdotcrit \simeq 2 \times 10^{13}$ \gpers ~which gives   $\Ld \simeq 8\times 10^{32}$ \ergpers, and  $\Lstar \simeq 4\times 10^{33}$ \ergpers~  during the transition. This seems to be consistent with the observations within the distance uncertainties of the source (Linares 2014). For IGR J18245--2452, $P = 3.93$ ms, $\Pdot < 1.3 \times 10^{-17}$ s s$^{-1}$, and $\Lx \lesssim 10^{32}$  \ergpers  ~in the RMSP state (Papitto et al. 2013).  For this source to be in the strong propeller phase with $\Lx = 10^{32}$  \ergpers, we estimate the lower limits $\mu_{26} > 1$ and $\Pdot > 3 \times 10^{-20}$ \spers. 

The tMSPs belong to the redback population of  MSPs  in compact LMXBs. Among the other redbacks,  $\Lx$,  $P$, and $\Pdot$ are known for 3 systems observed only in the RMSP states. These redbacks, with  properties similar to tMSPs in the RMSP state,  are thought to be strong tMSP candidates  (Linares 2014).  
For these sources, we have estimated $\mu_{26}$, $\Mdotin$, and $\Mdotcrit$  (Table 2).  The estimated $\Mdotcrit/ \Mdotin$ ratios are very close to unity for PSR J1723--2834 and  PSR J2215+5135, indicating that these sources are indeed good candidates for transition to the accretion phase with a few times increase in  $\Mdotin$. For PSR J2339--0533, we estimate that $\Mdotin$ is about 7 times lower than $\Mdotcrit$.

\section{ CONCLUSIONS}

Using the inner disk radius, $\rin$, estimated earlier for the propeller phase by Ertan (2017), we have modelled the torque acting on J1023 in the RMSP and LMXB states.  
The X-ray luminosity, $\Lx$, in the high mode of LMXB state is  produced by mass accretion onto the star  in the weak-propeller phase,  while the $\Lx$ in the  RMSP state is explained  by the emission from the inner disk in the strong-propeller phase. 

When the source is in the strong-propeller state, $\Mdotin \sim 2.8 \times 10^{12}$ \gpers~ the disk produces $\Lx \sim 10^{32}$ \ergpers, a small fraction of the orbitally modulated total $\Lx$.  A 6-fold increase in $\Mdotin$  takes the source to the weak-propeller phase with an accretion luminosity from the star's surface explaining the high $\Lx$. The change in $\rin$ between the high X-ray mode in the LMXB state (weak-propeller) and the RMXB state (strong propeller)  is only $\sim$ 9 \%   because of the very weak $\Mdotin$ dependence of $\rin$. The onset of accretion produces a negligible spin-up torque in comparison with the spin-down torque in the weak-propeller phase. The small decrease in $\rin$  increases the magnitude of the spin-down torque by $\sim$ 30 \%, which is in good agreement with the observations (Table 1). 
 
Occasional transitions between the strong-propeller and the weak-propeller phases when $\rin \simeq \rco$ could cause the observed transitions between the low X-ray mode and the high X-ray mode. With similar $\Mdotin$, mass accretion can produce $\Lx$ in the high mode, and when accretion is hindered, $\Ld$ with  $\rin \simeq \rco$ can  explain $\Lx$ in the low mode. In the LMXB state,  $\Mdotin$  is very close to $\Mdotcrit$ for  J1023. This could be the reason for occasional transitions between the high and low modes in the LMXB state.  

The scale $\rin \lesssim \rco$ and the weak $\Mdotin$  dependence of $\rin$ in our propeller model thus explain the torques and luminosities $\Lx$ in the different states. In the RMSP state the spin-down torque is supplied by the disk, as in the LMXB state, and not by dipole radiation. Radio pulsar activity proceeds with the disk until mass accretion disrupts it in either of the modes in the LMXB state.  

Finally, we have found that the redbacks PSR J1723--2834 and  PSR J2215+5135, which are observed in RMSP state, are strong tMSP candidates, close to their propeller-accretion transition rates (Table 2).  These sources could show transition to the LMXB  phase with small increases in their $\Mdotin$.


\section*{Acknowledgements}

We acknowledge research support from
T\"{U}B{\.I}TAK (The Scientific and Technological Research Council of
Turkey) through grant 117F144 and from Sabanc\i\ University.  We thank Ali Alpar for useful comments on the manuscripts. 












\bsp	
\label{lastpage}
\end{document}